\documentclass[12pt]{article}
\usepackage{latexsym,epsfig}
\usepackage{hyperref}
\usepackage{color}
\usepackage{blkarray}

\begin{document}

\centerline{\bf \large What Exactly is Antimatter (Gravitationally Speaking)?}
\smallskip

\centerline{\bf \large A Second Scenario }
\bigskip

\centerline{S. Menary}
\smallskip

\centerline{Dept. of Physics \& Astronomy}
\centerline{York University, 4700 Keele Street}
\centerline{Toronto, ON M3J 1P3, Canada}
\bigskip

\begin{abstract}
In \cite{antiart} I investigated the consequences of regarding the mass-energy of the fundamental fermions (quarks and leptons) and the Intermediate Vector Bosons (e.g., photon) as matter, and the fundamental antifermions (antiquarks and antileptons) as antimatter within the context of an antigravity universe, one where matter and antimatter repel gravitationally. Here I consider an alternative scenario
in which the Intermediate Vector Bosons, which are neither particle nor antiparticle, are gravitationally attracted to both fundamental fermions and antifermions. This leads to a prediction for the free-fall acceleration of antihydrogen of $a_{\bar{H}}=(0.78^{+0.11}_{-0.08})g$ (and most certainly less than $g$) as well as quite different expectations for the free-fall accelerations of the $\mu^+$ and positronium from those derived in \cite{antiart}. The cosmology which results from the premise presented here is little different from the standard cosmology (i.e., the $\Lambda$CDM model). One significant deviation is that there would be an increased accelerated expansion in the early moments after the Big Bang due to the gravitational repulsion between the fundamental fermions and antifermions. 
\end{abstract}

Antigravity is the hypothesis that matter and antimatter repel gravitationally. Clearly, if one wants to explore the ramifications of antigravity, it is necessary to precisely define what is meant by the terms "matter" and "antimatter". In \cite{antiart} I reviewed the various implementations of antigravity along with the arguments and counterarguments for and against them. I then investigated the scenario where the mass-energy of the fundamental fermions (quarks and leptons) and the Intermediate Vector Bosons (e.g., photon) constitute matter while the fundamental antifermions (antiquarks and antileptons) are antimatter.
Further, under Villata's CPT gravity\cite{villata}\footnote{Which is just GR but where Villata posited that if GR is invariant under $CPT$ transformations then matter and antimatter necessarily repel each other.}, this definition of matter and antimatter specifically led to the prediction that the free-fall acceleration of antihydrogen would be $a_{\bar{H}}=(0.33^{+0.23}_{-0.11})g$. The ALPHA-g result\cite{ALPHA-g} of $a_{\bar{H}}=(0.75\pm 0.13~({\rm stat.+syst.})\pm 0.16~({\rm simulation}))g$ is certainly consistent, at the $1\sigma$ level, with the standard gravity (GR) value of $a_{\bar{H}}=g$ but is also $\sim $1$\sigma$ from this prediction. 

In this article I want to explore a different interpretation of antigravity, one in which the fundamental fermions and antifermions would still repel each other gravitationally (i.e., still constitute matter and antimatter) but now where the intermediate vector bosons (IVB) are attracted to both fermions and antifermions. In keeping with the spirit of the title of this article, this now means the IVB no longer act as matter or antimatter and so I will coin the term "neumatter" when referring to their gravitational interactions from here on in.

\section{General Relativity, Newtonian Gravity, Composite Objects, and Antigravity}

The Einstein field equations without a cosmological constant term are:
$$R_{\mu\nu}-\frac{1}{2}g_{\mu\nu}R=8\pi GT_{\mu\nu}$$
where $T_{}\mu\nu$ is the energy-momentum tensor (EMT) of matter fields in the space-time, $G$ is Newton’s
constant, and $R_{\mu\nu}$ and $R=g^{\mu\nu}R_{\mu\nu}$ are the Ricci tensor and scalar curvature, respectively.
This can also be written in this way;
\begin{equation}
    G_{\mu\nu}=\frac{8\pi G}{c^4}T_{\mu\nu}
\end{equation}
where $G_{\mu\nu}$ is the Einstein tensor.

In \cite{antiart} I reviewed the partitioning of the EMT of the proton and the Lattice QCD results for the various contributions to the proton mass. Specifically, Ji found in 1995\cite{Ji1} that the total QCD EMT for the proton can be decomposed as
$T^{\mu\nu}=T_q^{\mu\nu}+T_g^{\mu\nu}+\hat{T}^{\mu\nu}$
where the last term is the "trace term" (a completely QFT effect). $T^{\mu\nu}$ and $\hat{T}^{\mu\nu}$ are scheme and scale ($\mu$) independent as is the sum $T_q^{\mu\nu}(\mu)+T_g^{\mu\nu}(\mu)$. One then has $M=M_q+M_g+M_a$ and, as stated in \cite{Ji2}, "If the quark masses are non-zero, one ends up with a decomposition with four terms, each of which are related to experimental observables and calculable in lattice QCD." A full lattice QCD calculation\cite{yang}(2018) gave the following 4-term (\'a la Ji) proton mass decomposition: quark condensate ($\sim$9\%), quark kinetic energy ($\sim$32\%), gluonic field strength ($\sim$37\%), and anomalous gluonic contribution ($\sim$23\%). The result was given for a specific choice of scale, $\mu$ = 2 GeV. The sum of quark kinetic energy and gluon field strength contributions is unaffected by the choice of scale but the relative sizes of them is scale dependent. From this we see that the bulk of the proton mass ($\sim$68\%) is due to the gluon (colour) field. The quark masses\cite{pdg} are $m_u$ = $2.16^{+0.49}_{-0.28}$ MeV and $m_d$ = $4.67^{+0.48}_{-0.17}$ MeV and so contribute $\sim$1\% to the proton's mass.

Due to the $CPT$ invariance of QCD, all of the EMT terms for the antiproton (as well as the antiquark masses) have the same numerical value as those for the proton with the only difference being that 
the quark kinetic energy term becomes the antiquark kinetic energy term and should carry the label antimatter. The quark condensate EMT term, essentially the "sea" quarks or virtual $q\bar{q}$ pairs generated by the colour field, is the same for the antiproton and proton and thus operates under gravity like matter. Thus, if we define $f$ and $\bar{f}$, the fractions of the object's EMT which are matter and antimatter, respectively, then we would have;
\begin{eqnarray}
    T^{\mu\nu}_{\bar{p}}&=&T^{\mu\nu}_{\bar{p}}(f+\bar{f})\\
    fT^{\mu\nu}_{\bar{p}}&=&T_{\bar{q}}^{\mu\nu}\nonumber\\
    \bar{f}T^{\mu\nu}_{\bar{p}}&=&T_g^{\mu\nu}+\hat{T}^{\mu\nu}\nonumber
\end{eqnarray}
The labeling\footnote{In the process of researching for this article I came across the interesting papers\cite{KFL1}\cite{KFL2} of Keh-Fei Liu in which he discusses in some detail the decomposition of the proton EMT into gluon and quark components. While the exact fraction apportioned to each is somewhat scheme dependent, for my purposes it doesn't matter -- the only requirement is that such a decomposition is shown to be theoretically justified.} is of course only relevant if antigravity is active.

The simplest and cleanest version of antigravity is that of Villata who purported\cite{villata} that matter-antimatter repulsion naturally arises if GR is invariant under $CPT$ transformations. Specifically he showed that applying $CPT$ to the equation of motion (the geodesic equation) for matter-matter interactions led to the standard form:
\begin{equation}
    \frac{dx^\lambda}{d\tau^2}=-\frac{m_{(g)}}{m_{(i)}}\frac{dx^\mu}{d\tau}\Gamma^\lambda_{\mu\nu}\frac{dx^\nu}{d\tau}
\end{equation}
where he thought it "may be useful to keep the ratio
$m_{(g)}/m_{(i)}$ = 1 visible in the equation." That is, the Equivalence Principle is still enforced. For matter-antimatter interactions he found:
\begin{equation}
    \frac{dx^\lambda}{d\tau^2}=-\frac{-m_{(g)}}{m_{(i)}}\frac{dx^\mu}{d\tau}\Gamma^\lambda_{\mu\nu}\frac{dx^\nu}{d\tau}
\end{equation}
where you can see that applying $CPT$ has introduced an extra minus sign, i.e., there is now matter-antimatter repulsion. 
He then noted that,
\begin{quote}
\begin{it}
    "The minus sign assigned to the gravitational mass
in eq. (4) must not be misinterpreted. It does not mean that $m_{(g)}$ has become negative,
since, according to our assumptions, i.e. $CPT$ invariance and weak equivalence principle, all
masses are and remain positive definite. As already said, the minus sign comes from the $PT$-
oddness of either $dx^\mu$ or $\Gamma^\lambda_{\mu\nu}$. Assigning it to the mass can just be useful for not losing it when dealing with the Newtonian approximation, where four-velocities disappear, together with
their changed signs. Similarly, the Newtonian-limit field $GM/r^2$ has lost the $PT$-oddness, so
that the minus sign of an antimatter field may consequently be assigned to $M$. As a result, we
would obtain the generalized Newton law 
\begin{equation}
F(r)=-G\frac{(\pm m)(\pm M )}{r^2} = \mp G\frac{mM}{r^2}
\end{equation}
where the minus sign refers to the gravitational self-attraction of both matter and antimatter,
while the plus sign indicates the gravitational repulsion between matter and antimatter."
\end{it}    
\end{quote}

For many physical situations (e.g., free-fall acceleration, galaxy rotation curves, etc.) it is not necessary to use the full GR but is sufficient to use Newtonian gravity. It is stated in \cite{khar} (after a rigorous derivation of Newton's Law of Gravitation from GR):
\begin{quote}
    \begin{it}
        The purpose of reviewing this textbook derivation was to show that in the weak gravitational field, non-relativistic, limit of gravity the distribution of mass and the distribution of the trace of the EMT are identical.
    \end{it}
\end{quote}
Thus there is a direct connection between the mass in Newton's Law of Gravitation and the EMT of GR. So, in \cite{antiart}, in order to implement eq. (5), I proposed the analogue of eq. (2) for mass;  
\begin{equation}
   M=m+\bar{m}=M(f+\bar{f}) 
\end{equation}
Newton's Law of Gravitation can then be written;
\begin{equation}
    F=-\frac{M_1\tilde{G}M_2}{r_{12}^2}
    =-M_1M_2\frac{(f_1+\bar{f}_1)\tilde{G}(f_2+\bar{f}_2)}{r_{12}^2}
\end{equation}
with $\tilde{G}=G$ for standard gravity while with antigravity we have;

$$\tilde{G}=G\tilde{U}=
G\left(\matrix{U_{ff}&U_{f\bar{f}}\cr
U_{\bar{f}f}&U_{\bar{f}\bar{f}}}\right)
=G\left(\matrix{+1&-1\cr 
-1&+1\cr}\right)$$

The fully expanded version of eq. (7) is then;
\begin{eqnarray}
    F&=&\frac{-GM_1M_2(f_1f_2+\bar{f}_1\bar{f}_2-f_1\bar{f}_2-\bar{f}_1f_2)}{r_{12}^2}\\
    &=&\frac{-GM_1M_2\left((1-\bar{f}_1)(1-\bar{f}_2)+\bar{f}_1\bar{f}_2-(1-\bar{f}_1)\bar{f}_2
    -\bar{f}_1(1-\bar{f}_2)\right)}{r_{12}^2}\\
    &=&\frac{-GM_1M_2}{r_{12}^2}(1+4\bar{f}_1\bar{f}_2-2\bar{f}_1-2\bar{f}_2)
\end{eqnarray}
Note that, as must be the case, for pure matter ($\bar{f}=0$) and antimatter ($\bar{f}=1$) states you have:
$$F_{ff}=F_{\bar{f}\bar{f}}=-F_{f\bar{f}}=-G\frac{M_1M_2}{r_{12}^2}$$

Finally, just to complete this recap of the relevant sections of \cite{antiart}, for the case of antihydrogen falling in the gravitational field of the Earth, eq. (10) becomes particularly simple since then $\bar{f}_1=0$. Equating $M_1=M_E$ and $r_{12}=R_E$ we then have:
$$  F=\frac{-GM_EM_{\bar{H}}(1-2\bar{f}_{\bar{H}})}{R_E^2} $$ 
and it follows that the free-fall acceleration of antihydrogen is:
$$a_{\bar{H}}=(1-2\bar{f}_{\bar{H}})g$$
For all intents and purposes, $\bar{f}_{\bar{H}}=\bar{f}_{\bar{p}}= 0.33^{+0.06}_{-0.10}$ since $m_e/m_p$ is $5.4\times 10^{-4}$ leading to the prediction for 
the free-fall acceleration of antihydrogen of $a_{\bar{H}}=(0.33^{+0.23}_{-0.11})g$.

\section{Incorporating Neumatter}

The original motivation in \cite{antiart} for assigning the label "matter" to the intermediate vector bosons (IVB) was the observation of gravitational lensing (i.e., the bending of light paths consistent with gravitational attraction) and precision tests of the Weak Equivalence Principle (WEP) which showed that binding energy (atomic, nuclear, and nucleonic) acts like matter under gravity. Also, Villata showed that the photon follows a geodesic consistent with it acting like matter\cite{vill2}. In principle, however, these observations (and Villata's calculation) do not preclude the possibility that the IVB are also attracted to objects composed of antimatter, i.e., that they don't act like "matter" under antigravity in the strictest sense since they could be attracted to both matter (e.g., quark mass-energy) and antimatter (e.g., antilepton mass-energy). 

In order to explore this scenario I will expand the types of matter to include neumatter (i.e., IVB) so that we now have three types of matter -- fermionic, antifermionic, and bosonic -- that can, in principle, attract or repel each other gravitationally. Equation (6) can more properly be understood as a relation between inertial and gravitational mass. Using the labels +, -, and 0 for fermionic matter (matter), antifermionic matter (antimatter), and bosonic matter (neumatter), respectively, eq. (6) becomes;
\begin{equation}
    M_G=M_I(f^++f^0+f^-)\equiv M(f^++f^0+f^-)=M\sum_if^i
\end{equation}
where $i$ = +, 0, and -. 
The equivalence of gravitational and inertial mass is manifestly satisfied here since $f^++f^0+f^-=1$.
The $\tilde{U}$ matrix becomes the $3\times 3$ $U^{ij}$ matrix;

\[U^{ij}=
\begin{blockarray}{cccc}
& f^+& f^0 & f^- \\
\begin{block}{c(ccc)}
  f^+ & U^{++}&U^{+0}&U^{+-}  \\
  f^0 & U^{0+}&U^{00}&U^{0-} \\
  f^- & U^{-+}&U^{-0}&U^{--} \\
\end{block}
\end{blockarray}
 \]

and the force equation, eq. (7), then becomes;
\begin{equation}
    F=-\frac{M_{G1}\tilde{G}M_{G2}}{r_{12}^2}
    =-G\frac{M_1M_2}{r_{12}^2}\left(\sum_i\sum_jf_1^iU^{ij}f_2^j\right)
\end{equation}

It's now possible to delineate the different scenarios via the elements of the $\tilde{U}$ matrix. For example, in standard gravity; 
$$U_S=\left(\matrix{+1&+1&+1\cr+1&+1&+1\cr+1&+1&+1
}\right)$$
That is, the gravitational interaction is independent of the type of matter involved so there is effectively only one type, matter, and $i=j=+$. Therefore, $M\sum f^i=M$ and $U_S\rightarrow U^{++}=1$ leading to the usual formula;
\begin{equation}
    F_S=-G\frac{M_1M_2}{r_{12}^2}
\end{equation}

For the antigravity scenario of \cite{antiart} you'd have;
$$U_b=\left(\matrix{+1&+1&-1\cr+1&+1&-1\cr-1&-1&+1
}\right)$$
In this case, fermionic and bosonic matter are equivalent under the gravitational interaction so there are effectively just two types -- matter and antimatter. We would then have $\sum f^i=f+\bar{f}$ and  $U^{ij}$ would collapse to;
$$U_b=\left(\matrix{+1&-1\cr-1&+1
}\right)$$
resulting in;
\begin{equation}
   F_b= \frac{-GM_1M_2}{r_{12}^2}\left(1+4\bar{f}_1\bar{f}_2-2\bar{f}_1-2\bar{f}_2\right)
     =F_S\left(1+4\bar{f}_1\bar{f}_2-2\bar{f}_1-2\bar{f}_2\right)
\end{equation}

For wont of a better term I'll call the previous scenario, where IVB and fermions act as matter and repel antifermions, "bosonic antigravity" while for this final scenario, where fermions and antifermions repel while bosonic matter (IVB) is attracted to both, I'll term it "fermionic antigravity." It is represented by;
$$U_f=\left(\matrix{+1&+1&-1\cr+1&+1&+1\cr-1&+1&+1
}\right)$$
This leads to;
\begin{equation}
    F_f=-\frac{GM_1M_2}{r_{12}^2}\left[1-2\left(f_1^+f_2^-+f_1^-f_2^+\right)\right]=F_S\left[1-2\left(f_1^+f_2^-+f_1^-f_2^+\right)\right]
\end{equation}
\subsection{Free-fall Acceleration}
One big difference between bosonic 
and fermionic antigravity is in the prediction for the free-fall acceleration of objects either wholly or partially composed of antifermions.  
For example, in fermionic antigravity, eq. (15), we would get for the free-fall acceleration of antihydrogen;
\begin{equation}
    a_{\bar{H}}=\left[1-2\left(f_{Earth}^+f_{\bar{H}}^-+f_{Earth}^-f_{\bar{H}}^+\right)\right]g
\end{equation}
The fermionic fraction of the Earth's mass is essentially the fermionic fraction of the proton and neutron masses (the electrons add a negligible amount) and assuming $f_n^+=f_p^+$, which is presumably also true to $\sim$1\%, we'd have $f_{Earth}^+\approx f_p^+$. Therefore,
$$f_{Earth}^+\approx f_p^+=f_{\bar{p}}^-=f_{\bar{H}}^-=0.33^{+0.06}_{-0.10}$$
and since $f_{Earth}^-=f_{\bar{H}}^+=0$, we arrive at the final result;
\begin{equation}
    a_{\bar{H}}=\left[1-2f_{Earth}^+f_{\bar{H}}^-\right]g=\left[1-2\left(f_p^+\right)^2\right]g=\left(0.78^{+0.11}_{-0.08}\right)g
\end{equation}
While this prediction does agree improbably well with the ALPHA\_g measurement, it is also many sigma\footnote{The $\chi^2$ function is certainly not symmetric around the minimum. This reflects the fact that the Lattice QCD prediction for the fraction of the antiproton mass due to the antiquark kinetic energy has a large $1\sigma$ uncertainty but it is many sigma from being zero.} from the standard gravity prediction of $a_{\bar{H}}=g$. A better statement would be that, in fermionic antigravity, $a_{\bar{H}}$ must be less than $g$.

As for the $\mu^+$, in bosonic antigravity
one would get $a_{\mu^+}=-g$. The $\mu^+$ is an antifermion so $f^-_{\mu^+}=1$ while $f^+_{\mu^+}=0$ and in fermionic antigravity we'd then have;
$$ a_{\mu^+}=\left[1-2f_{Earth}^+f^-_{\mu^+}\right]g=\left[1-2f_p^+\right]g=\left(0.33^{+0.23}_{-0.11}\right)g=a_{\bar{H}}$$
Since the gravitational force is so much smaller than the electric force it is difficult to measure the free-fall acceleration of the charged muon because of the need to precisely understand and control any stray electric fields. There is a proposal\cite{mage} to measure the free-fall acceleration of antimuonium, the bound state of an electron and an antimuon. The binding energy of antimuonium (essentially that of hydrogen, i.e., 13.6 eV) and the kinetic energy of the electron are negligible compared to the $\mu^+$ mass of 105.66 MeV = 206.77$m_e$. So for antimuonium:
$$\bar{f}_{\rm Mu}=\frac{m_\mu +K_\mu}{m_\mu+m_e-E_B}\approx \frac{m_\mu}{m_\mu+m_e}=\frac{206.77m_e}{207.66m_e+me}=0.995$$
and in \cite{antiart} it was found that;
$$a_{\rm Mu}=(1-2\bar{f}_{\rm Mu})g=(1-2(0.995))g=-0.99g$$
For fermionic antigravity we would have;
$$a_{\rm Mu}=\left[1-2f_{Earth}^+f^-_{\rm Mu}\right]g=\left[1-2f_p^+(0.995)\right]g=\left(0.34^{+0.23}_{-0.11}\right)g$$

Finally, for positronium, where $f_P^-=f_P^+=1/2$ (the binding energy of 7 eV is insignificant), it was found that $a_P\approx 0g$ in \cite{antiart} whereas for fermionic antigravity you'd have;
   $$ a_P=\left[1-2f_{Earth}^+f^-_P\right]g=\left[1-f_p^+\right]g=\left(0.67^{+0.10}_{-0.06}\right)g$$

\begin{center}
\begin{table}[h!]
\begin{tabular}{|c||c|c|c|}
\hline
Quantity & Standard Gravity&Bosonic Antigravity&Fermionic Antigravity\\
\hline \hline
$a_{\bar{H}}$&$g$ & $\left(0.33^{+0.23}_{-0.11}\right)g$ & $\left(0.78^{+0.11}_{-0.08}\right)g$\\
\hline
$a_{\rm Mu}$ & $g$ & ~~~~~~~-0.99~$g$ & $\left(0.34^{+0.23}_{-0.11}\right)g$\\
\hline
$a_P$ & $g$ & ~~~~~~~~~~~0~$g$ & $\left(0.67^{+0.10}_{-0.06}\right)g$ \\
\hline
\end{tabular}
\caption{Comparison of the predictions from Standard Gravity, Bosonic Antigravity\cite{antiart}, and Fermionic Antigravity for the free-fall acceleration of antihydrogen, antimuonium, and positronium. }
\end{table}
\end{center}

\subsection{Cosmological Implications}

The antigravity scenario of \cite{antiart}, eq. (14), leads to a rich cosmological phenomenology as described in some detail in \cite{anticosmos}. This was primarily due to the fact that
$\bar{f}\approx$ 1/3 for antihydrogen and $\bar{f}$ = 0 for hydrogen results in;
\begin{equation}
F_b^{\bar{H}\bar{H}}=\left(\frac{1}{3}\right)F_b^{\bar{H}H}=\left(\frac{1}{9}\right)F_b^{HH}
\end{equation}
The same relations are true for antihelium and helium assuming $M_{\bar{n}}=M_n$, as required by the $CPT$ invariance of QCD. So even though there would be equal amounts of hydrogen and antihydrogen, this resulted in there being a paucity of antistars, as well as the formation of cosmic voids, an effective modified gravity explanation for galactic rotation curves, variable accelerated expansion, etc. 

For fermionic antigravity, eq. (15), we would have;
\begin{equation}
F_f^{\bar{H}\bar{H}}=\left(\frac{7}{9}\right)F_f^{\bar{H}H}=F_f^{HH}
\end{equation}
That is, the force between antihydrogen atoms is the same as that between hydrogen atoms so if there were equal amounts of antihydrogen and hydrogen then you would expect equal numbers of antistars as stars, a situation disfavoured by observation (e.g., see \cite{rujula}). This is not a problem if one assumes the same mechanism which removes the antifermions early in the universe's evolution\footnote{This can be done via, for example, the famous Sakharov conditions\cite{sakharov}.} as is at work in the standard cosmology. 

Also, in \cite{antiart} we had $F_b^{\gamma \bar{f}}=-F_b^{\gamma f}$
so gravitational lensing was completely altered depending on whether photons encountered matter or antimatter along their path since photons are actually {\it de-focused} by regions occupied by, for example, antineutrinos.   
In fermionic antigravity $F_f^{\gamma \bar{f}}=F_f^{\gamma f}$
meaning there would be no affect on gravitational lensing due to the presence of antimatter in bosonic antigravity.

One is left, therefore, with a fermionic antigravity universe very much like that described by the currently favoured $\Lambda$CDM model. The sole difference is that one would expect a greater acceleration (and perhaps a different Hubble constant) from that of the standard cosmology in the very early universe, i.e., before the generation of the baryon asymmetry, since there would be the gravitational repulsion between the fundamental fermions and antifermions. After this point it doesn't matter whether there is some form of antigravity acting or not as there is, like in the standard cosmology, very little antimatter for it to act on.

\section{Conclusions}
In a previous paper\cite{antiart} I explored what happens if you define the mass-energy of the fundamental fermions (quarks and leptons) and intermediate vector bosons (photon, $W^-$, etc.) as matter while the mass-energy of the fundamental antifermions (antiquarks and antileptons) is antimatter. In this paper I examined the ramifications still having the mass-energy of the fundamental fermions and antifermions repelling gravitationally but having the IVB attracted gravitationally to both the fermions and antifermions. I found that this leads to the requirement that $a_{\bar{H}}$ will be some 5\% to 20\% less than $g$. Furthermore, 
the universe in this scenario would look essentially just like that described by the present $\Lambda$CDM model. The only evidence for the scenario presented here would be in the free-fall acceleration of antihydrogen and antimuonium and perhaps also in a greater acceleration (stronger Dark Energy) in the very early universe.

\end{document}